\documentclass[3p,times]{elsarticle}

\usepackage{ecrc}

\volume{00}

\firstpage{1}

\journalname{Physics Procedia}

\runauth{B. E. Sauer et al.}

\jid{phpro}

\jnltitlelogo{Physics Procedia}

\usepackage{amssymb}

\usepackage[figuresright]{rotating}

\newcommand{\Eeff}{E_\textit{\footnotesize eff}}

\begin{document}

\begin{frontmatter}



\dochead{Physics of Fundamental Symmetries and Interactions -- PSI2010}

\title{Prospects for the measurement of the electron electric dipole moment using YbF}


\author{B. E. Sauer} \author{J. J. Hudson} \author{D. M. Kara} \author{I. J. Smallman} \author{M. R. Tarbutt} \author{E. A. Hinds}

\address{Centre for Cold Matter, Quantum Optics and Laser Science, Blackett Laboratory,\\
Imperial College London, Prince Consort Road,
London SW7 2AZ, United Kingdom}

\begin{abstract}
We discuss an experiment underway at Imperial College London to measure the permanent electric dipole moment (EDM) of the electron using a molecular beam of YbF. We describe the measurement method, which uses a combination of laser and radiofrequency resonance techniques to detect the spin precession of the YbF molecule in a strong electric field. We pay particular attention to the analysis scheme and explore some of the possible systematic effects which might mimic the EDM signal. Finally, we describe technical improvements which should increase the sensitivity by more than an order of magnitude over the current experimental limit.
\end{abstract}


\end{frontmatter}


\section{Introduction}

While in the standard model  the electric dipole moment (EDM, or $d_e$) of the electron is very nearly zero,  many extensions to the standard model naturally predict values~\cite{Khriplovich97,Commins99,Commins10} close to the current experimental limit. This limit,  $d_e =(6.9\pm7.4)\times 10^{-28}\,$e.cm, was obtained in an experiment using atomic thallium \cite{Regan02}. Since of the  electron EDM predicted by the standard model is very much smaler, the current experiment is specifically sensitive to physics beyond the standard model. The existing experimental result already limits possible new physics, for example by excluding some supersymmetric models with large CP violating phases.  However, further improvement in the measurement sensitivity is needed to test predictions of $d_e$ in the $10^{-28}$ to $10^{-29}$\,e.cm range.

Atomic and molecular electron EDM experiments search for a spin dependent interaction $\hat{\sigma}\cdot \vec{E}$, where $\hat{\sigma}$ is the electron spin and $\vec{E}$ is an applied external electric field. Sandars \cite{Sandars65} was the first to note that the interaction can be amplified in a heavy atom. This amplification can be characterized by  an effective field $\Eeff$ which can be much larger than the applied field. A few years after Sandars discovered the atomic enhancement he also pointed out \cite{Sandars67} that heavy polar molecules, being far more polarizable than atoms, have a great advantage over atoms in that $\Eeff$  can saturate at a very large value in a modest laboratory field. Over the past decade therefore we have developed an EDM experiment~\cite{Hudson02} using the polar molecule YbF. This molecule contains a heavy nucleus and has the great advantage that it is straightforward to produce and detect. In our experiment on YbF the interaction energy due to $d_e$ is 220 times larger than Ref.~\cite{Regan02} obtained using Tl atoms and a much higher electric field \cite{Hinds97,enhancement}.  The YbF experiment has very different sensitivity to systematic effects than the atomic experiment. In particular, the motional magnetic field, which limited the Tl measurement, generates a negligible systematic error in the case of YbF \cite{Sauer01}. For these reasons, YbF is able to offer a substantial improvement in sensitivity over the Tl experiment, even though the overall beam intensity is much smaller in the molecular experiment.

We note that atomic and molecular systems are also sensitive to other interactions that violate time reversal symmetry  as well as to the electron EDM \cite{Pospelov10}. As this sensitivity can differ for different systems, it is worthwhile to make measurements on a variety of atomic and molecular species, even at similar levels of sensitivity. In this paper, we discuss the current measurement and data analysis techniques using YbF, and also the expected progress from technical improvements.

\section{Method}

\begin{figure}[t]
\begin{center}
\includegraphics[height=8.0cm]{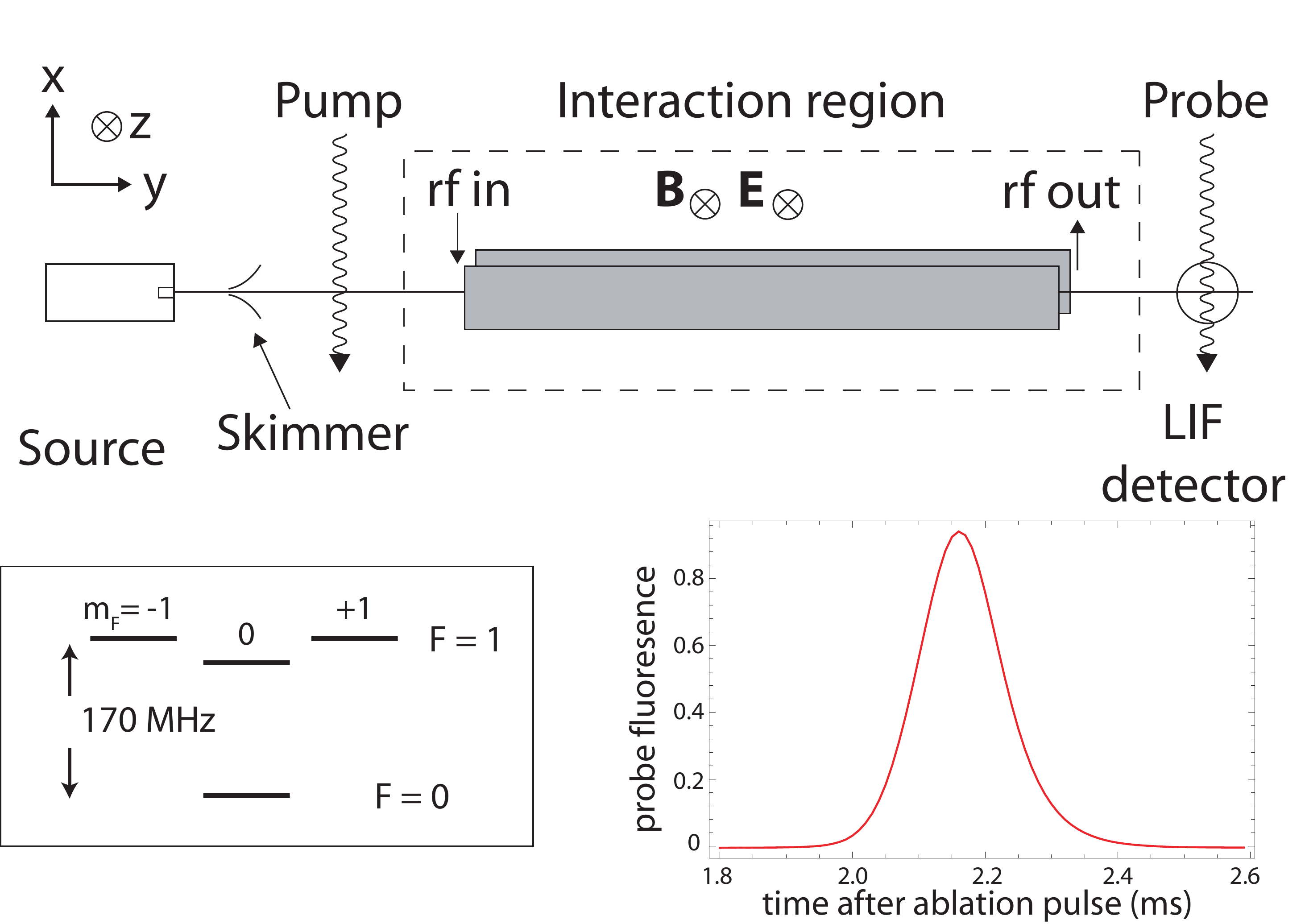}
\end{center}
\caption{\label{fig:apparatus}Schematic diagram of the molecular beam apparatus. Lower right: typical beam pulse measured via laser induced fluoresence (LIF) on the probe photomultiplier. The data shown are averaged over a 4096 individual beam pulses. Box: Ground-state hyperfine structure of the YbF molecule, showing the tensor Stark splitting of the $F=1$ state in an electric field.}
\end{figure}

Figure\,\ref{fig:apparatus} shows the essential features of the apparatus \cite{Hudson07}. YbF molecules are formed by laser ablation in a supersonic expansion \cite{Tarbutt02}. The YbF molecules emerge as gas pulses with a temperature of about 1\,K and a forward velocity of about 600\,m/s. This low temperature means that the population is very nearly all in the electronic and vibrational ground state $X^2 \Sigma^+(v=0)$, with a significant thermal population in the ground rotational state. The lowest rotational state has two hyperfine levels F=0 and F=1, separated by about 170\,MHz \cite{Sauer96}, as shown inset in Fig.\,\ref{fig:apparatus}. These are equally populated in the source. The hyperfine levels are easily resolved using a single-mode CW dye laser running near 552\,nm. This laser beam is split to provide  a frequency-shifted pump beam and a probe beam: the pump empties out the F=1 population by exciting the Q(0) F=1 line of the A-X transition, while the probe is used for fluorescence detection of the F=0 molecules.  The resulting photomultiplier signal is digitized, producing time-of-flight signals such as that shown in Fig.\,\ref{fig:apparatus}.

The experiment measures the spin precession of the electron in an electric field. After the molecular pulse passes the pump laser, it enters a magnetically shielded interaction region containing a pair of electric field plates. In the interaction region we apply static electric and magnetic fields $(E, B)$, typically with $E=\pm10\,$kV/cm and $B=\pm13\,$nT. The YbF molecules are strongly polarized along the E field, which therefore defines the z-axis. The electric field plate structure doubles as a transmission line to co- or counter-propagate 170\,MHz radiation along the beam direction. When the packet of F=0 molecules is well inside the plates, a short rf $\pi$-pulse creates the  $F=1$ superposition $|x\rangle =\frac {1}{\sqrt{2}} \left(|m_F=+1\rangle+|m_F=-1\rangle\right)$. We can ignore the $|F=1, m_F=0\rangle$ state because the tensor Stark shift moves it well out of resonance. The molecules then evolve freely for a time $T$, experiencing the Zeeman shift $+\mu_B m_FB$ of the $m_F=\pm 1$ states \cite{Steimle09} and the EDM interaction given by $-d_e\Eeff\,m_F$. If the molecules were fully polarized, $\Eeff$ would be 26\,GV/cm  \cite{enhancement}, however in our applied field the polarization only reaches 50\% \cite{Sauer96}, so $\Eeff=13$\,GV/cm. The two $m_F$ components develop a relative phase shift of $2\phi=2(- d_e \Eeff + \mu_B B)T/\hbar $, which causes a rotation from the state $|x\rangle$ towards the state $|y\rangle =\frac {1}{\sqrt{2}}\left(|+1\rangle-|-1\rangle\right)$. A second $\pi$-pulse is then applied, resulting in a final F=0 population proportional to $\cos^2\phi$. The probe laser measures this population via laser induced fluorescence. For non-ideal rf pulses the exact lineshape departs slightly from the simple $\cos^2\phi$ form, as is discussed in \cite{Tarbutt09a}. The overall sensitivity of the experiment increases linearly with the interaction time $T$ and with the square root of the molecular beam intensity.

The sensitivity to small changes in rotation angle is maximized by tuning the applied magnetic field to operate at $\phi=\pm\pi/4$, the maximum slope of the $\cos^2\phi$ signal. Reversing the applied electric field produces a small phase shift $\delta \phi=2 d_e \Eeff T/\hbar$, leading to a change in the detected fluoresence.  The slope of the  $\cos^2\phi$ signal at $\phi=\pm\pi/4$ can be calibrated by making a small known step in magnetic field magnitude. The simplest experiment therefore has three switches, the signs of E and B and the step in the magnitude of B. Successive beam pulses are measured in different switch states. In reality, we switch nine parameters, giving $2^9=512$ different machine states. The additional switch parameters are the laser frequency and the rf pulse amplitudes, frequencies and relative phase. We group the shots into `blocks' of 4096 beam pulses, over which all combinations of switch states are covered equally eight times. If for the moment we ignore the small effect of the switched parameters on the beam intensity, then the difference between fluorescence signals with $E$ parallel and antiparallel to $B$ determines $d_e$. In addition, this same dataset gives information on other correlations. For example the $B$ correlation by itself measures how well the operating fields switch exactly between $\phi=\pm\pi/4$. This provides an error signal at the end of each block that is fed back to compensate for small drifts of the ambient field. At the end of each block the data are analysed and the results are used to feed back to the rf generator and laser in order to keep the resonances maximized. Also after every block, the linear polarization of the pump and probe beams is randomly rotated. This is precaution relating to the very weak background of non-resonant fluorescence from the F=1 molecules, which ensures that the fluorescence has no sensitivity to the F=1 polarisation in the $xy$ plane. Each reversal of the electric field direction results in about 14s of deadtime while the switching transients die away. Including this, a block of data takes approximately six minutes to accumulate.

Manual reversals of the connections between the switching apparatus and interaction region are important because they provide an additional way to check for systematic errors, independent of the computer controlled switches. There are three manual reversals: $E$, $B$ and the direction of rf propagation along the field plates. These manual changes are made infrequently - typically one switch per day. Over the course of a month-long data run we balance the number of blocks taken in all eight of the manual states.

The beam line has one layer of magnetic shielding inside the vacuum system and a second outside. A fluxgate magnetometer between the shields measures the magnetic field parallel to $E$ near the center of the interaction region. A number of other magnetometers are used to check that the switching hardware does not inadvertently create magnetic fields. We double all of the logical outputs from the control system so that every high output is balanced by a low output. This makes the power supply current to the switching electronics independent of machine state. The data aquisition system also monitors two null voltages, a battery and a short circuit. These are used to check for noise and systematic errors arising from the switching and signal processing electronics, and also provide null data as input to the analysis routines.

\section{Analysis}

To derive measurements from a  block of data, we calculate how the detected signal is correlated with each of the 512 combinations of automatic switch reversals.  The data analysis routinue then adds a blind offset to the EDM which is not removed until all aspects of the analysis are completed. The discussion in this paper is based on a preliminary analysis of more than 8000 blocks of data with the electric field set to $E=10\,$kV/cm.   The uncertainty in the mean is found using the bootstrap method \cite{Efron86} to determine the symmetric confidence level \cite{Efron87}. This takes into account the slightly non-normal distribution of the EDM values. These measurements give a raw uncertainty for the electron EDM of about $d_e=\pm 6\times 10^{-28}\,$e.cm at the 68\% confidence level.

Imperfections in the reversals make the analysis more complicated. For example, a small change in the magnitude of E when it reverses detunes the rf transitions through the Stark effect, leading to an intensity change correlated with the direction of E. By itself this does not contribute to the EDM signal, however it can combine with an offset in magnetic field to produce a change in signal correlated with the relative directions of E and B. This is easily corrected using the E and B correlations extracted from the block data \cite{Cho91}. More generally, we examine all 511 possible correlations other than the $E\cdot B$ correlation to ensure that the experiment operates correctly and the known corrections are applied according to the individual parameter values. In addition, the blocks can be grouped according to the manual-reversal state of the electric and/or magnetic field to provide an independent check for any false asymmetry generated by the automatic switching.  Thus far we find no significant difference in any of the channels between automatic and manual switching.

\section{Systematic tests}

A very large number of possible systematic effects are ruled out by the combination of the null channels and the manual state reversals. One effect we do find is a correlation of the molecular signal with the magnetometer signal measured outside the inner shield, which shows (unsurprisingly) that the $z$ component of the magnetic field variations inside and outside the shield are correlated. This correlation has a negligible effect on the central value of the EDM, but correcting for it slightly narrows the block by block distribution, thus slightly reducing the uncertainty in $d_e$.

\begin{figure}[t]
\begin{center}
\includegraphics[height=6cm]{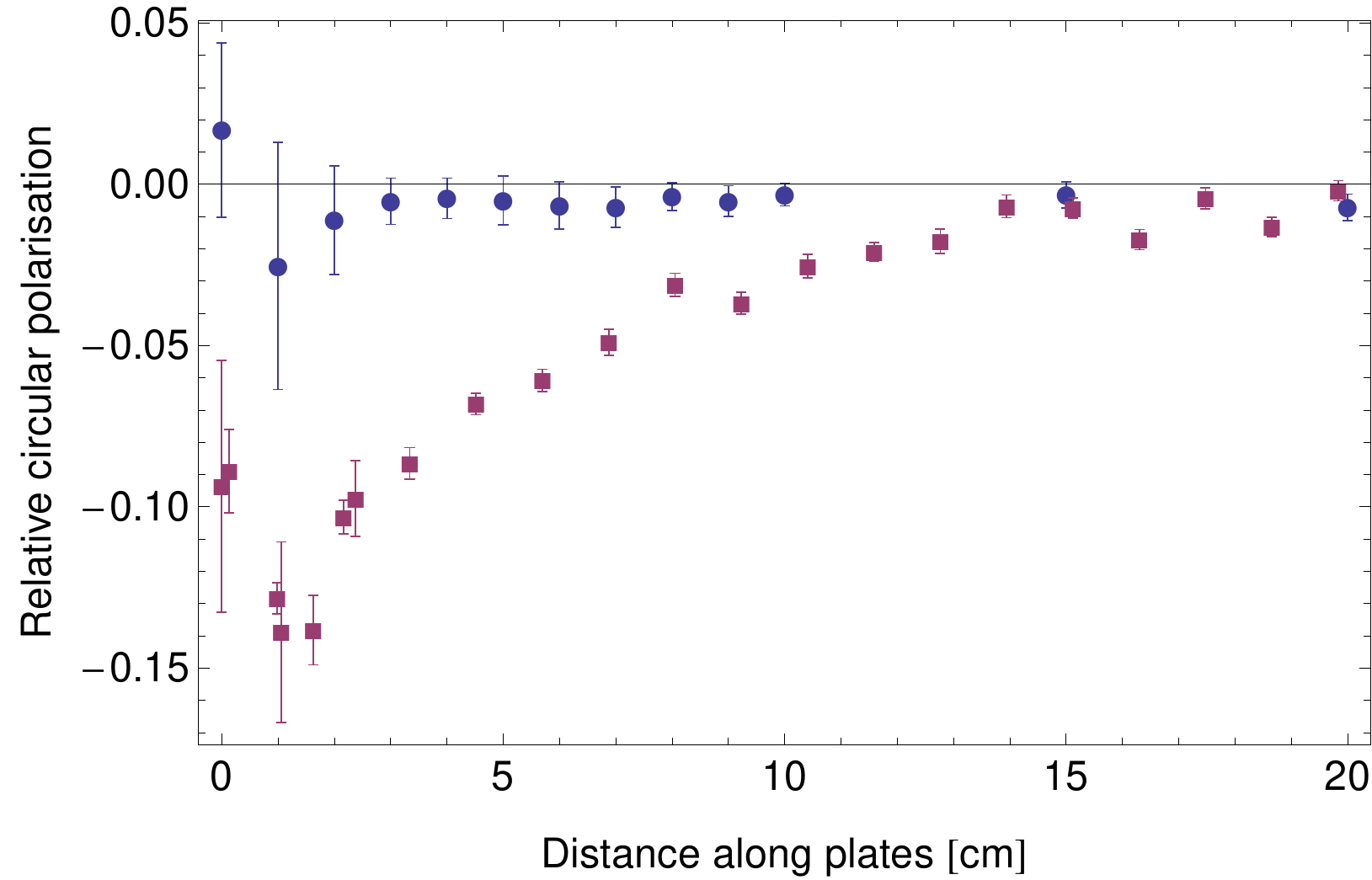}
\end{center}
\caption{\label{fig:polar}Purple squares: fractional circular polarization of rf field as a function of distance along the electric field plates, as measured in the exisiting apparatus using the YbF molecules as probes. Blue circles: fractional circular polarization measured using rf probes in a test apparatus with an improved rf coupling scheme. The usable length of the plates will be greatly increased by the new coupling scheme.}
\end{figure}

In auxiliary experiments we use the YbF molecules themselves to map out the spatial variation of the electric, magnetic, and rf fields \cite{Hudson07}. Imperfections in these fields can also potentially produce a false EDM signal. The rf field is nominally linearly polarized. However, at each end of the plates it has a few cm of transient ellipticity, as shown in figure \ref{fig:polar}, due to the way the coaxial feed connects to the plates. This puts more amplitude into one one of the $|m_F|=1$ states than the other, so that the rf pulse does not produce a perfect $|x\rangle$ state. While this would not in itself produce any systematic error, we nonetheless avoid it by choosing the rf pulse timings such that the molecules are well inside the electric field plates when they make their rf transitions. This has the disadvantage of reducing the coherence time $T$ during which the EDM phase accumulates.

Two imperfections in the electric field reversal contribute to systematic effects: i) its magnitude can change when it reverses and, ii) its symmetry about ground potential can change, if the individual positive and negative power supplies do not produce exactly the same magnitude of output with respect to ground. The Stark shift of the YbF hyperfine transition makes case i) evident in the data analysis as a correlation of the rf frequency that maximizes the signal with the electric field direction. We find the magnitude of the electric field asymmetry to be about $150\pm10\,$mV/cm. The imperfection in case ii) rotates the polarization axis of the molecules when the field direction reverses. This can combine with other imperfections  to produce a false EDM.  Both of these effects are investigated by recording EDM data while exaggerating the particular imperfection. A correction is then applied to the EDM value obtained from the data taken when the imperfections are small. The total uncertainty due to these systematic corrections in our analysis is at most at the $2\times 10^{-28}$\,e.cm level, negligible compared to our statistical uncertainty. However, it does require a significant experimental effort to exaggerate the effects, measure them, and demonstrate they do not shift the central value of the EDM. There are other possible sources of systematic error, for example leakage currents or the geometric phase \cite{Tarbutt09b}, but their effect in our apparatus is even smaller.

\section{Future improvements}

Our effort to improve the sensitivity of the experiment has three strands: to increase the spin precession time $T$, to increase the signal intensity, and to reduce the size of systematic effects which require empirical measurement. It will be very straightforward to increase $T$ by improving the rf coupling to the electric field plates so that the rf transitions can be driven closer to the ends of the plates. Figure \ref{fig:polar} displays the results from measurements on a set of  test plates with symmetric rf coupling. The improved scheme  vastly reduces the polarization ellipticity.  This will increase the usable plate length by about 40\%, with a corresponding increase in sensitivity once we have installed this in the EDM apparatus. We have also built a new 552\,nm laser system consisting of an infrared diode laser and a fiber amplifier. The infrared light is doubled to 552\,nm in a periodically poled doubling crystal. This system should be much more stable and reliable than the current ring dye laser. We will use light from this laser to add a second pump beam, transferring population from the bystander $N=2$ rotational level to the $N=0$ level used in the EDM experiment \cite{Hudson02}. Together these two improvements will allow us to measure at the $2\times 10^{-28}$\,e.cm level. At this sensitivity we will be limited by laboratory magnetic field noise and will need to add a third layer of magnetic shielding.

We will also improve the interaction with the electric field. We have designed and built optically coupled, microprocessor based electric field controllers which will allow very fine control of the high voltage power supplies. In addition, these open the possibility of active feedback of the electric field. This should drastically reduce the impact of possible systematics related to the asymmetry of the applied electric field. Electric and magnetic field gradients are also a concern, as they unavoidably introduce imperfections in the $\pi$-pulses by detuning the molecular transitions over the finite size of the YbF beam pulse. Our current electric field plates are slightly warped from the pressure of their high voltage feedthroughs, a situation which will be  straightforward to correct. We will upgrade our rf system using a 1\,kW pulse amplifier with a directly synthesized source. This will allow us to use much shorter $\pi$-pulses, which will reduce the distance the molecules move during the pulse and also broaden the molecular transitions. Both of these effects will reduce the sensitivity to field gradients along the beam path.

In parallel with these technical improvements, we have developed a new source of cold, slow YbF molecules. This is a buffer gas source \cite{Skoff09, Skoff10}, in which YbF molecules are produced by laser ablation of Yb and $\rm{AlF}_3$ into a cell of 4\,K helium gas. These precursors react to produce YbF, which thermalizes with the He and then escapes the cell as a slow, intense beam. We have demonstrated beams with a velocity of $200\,\rm{m/s}$, rather than the $600\,\rm{m/s}$ of our current source, and with a substantially ($10\times)$ higher intensity. A source of this type promises to make the EDM experiment more sensitive by an order of magnitude - a very exciting prospect, as it may well allow clear observation of a nonzero permanent EDM and the associated physics beyond the standard model.


\section{Acknowledgments}

We acknowledge laser development work by Suresh Doravari and expert technical assistance from Jon Dyne. This work was supported by the UK research councils STFC and EPSRC, and by the Royal Society.








\end{document}